\documentstyle[preprint,floats,tighten,prl,aps]{revtex}
\def \pom {{\scriptscriptstyle \kern -0.1em I \kern -0.25em P}}
\input epsf.tex
\def\desepsf(#1 width #2){\epsfxsize=#2 \epsfbox{#1}}
\begin{document}
\preprint{\vbox{
\hbox{CERN-TH/98-168} 
\hbox{ETH-TH/98-14} 
\hbox{OITS 648}
\hbox{June 1998}}  }
\draft

\title{Diffractive deeply inelastic scattering of hadronic states \\ 
with small transverse size }
\author{F.\ Hautmann$^{a}$, Z.\ Kunszt$^{b}$ and D.E.\ Soper$^{a}$}
\address{$^{a}$ Institute of Theoretical Science, 
University of Oregon, Eugene OR 97403, USA}
\address{$^{b}$ Institute of Theoretical Physics, 
ETH, CH-8093 Z\"urich, Switzerland}

\maketitle

\begin{abstract}
Diffractive deeply inelastic scattering from a hadron is 
described in terms of diffractive quark and gluon distributions. 
If the transverse size of the hadronic state is sufficiently small, 
these distributions are calculable using perturbation theory. We 
present such a calculation and   discuss the 
underlying dynamics.  We 
comment on the relation between this dynamics and the 
pattern of scaling violation observed in the hard diffraction 
of large-size states at HERA.  
\end{abstract} 



The data~\cite{origdata,partanal} from HERA on 
diffractive deeply inelastic scattering 
can be interpreted~\cite{partanal,gest}  
in terms of diffractive 
parton distribution functions, 
\begin{equation}
\label{pdf}
{ df_{a/A}^{\rm diff}(\xi,x_\pom,t,\mu)
\over dx_\pom\, dt} \hspace*{0.3 cm} .
\end{equation}
This function gives the 
joint 
probability per unit $d\xi$ to find a parton of 
type $a$ in a hadron of type $A$ 
and to find that 
 the hadron is diffractively scattered. 
Specifically, the hadron appears in the final state having lost a fraction 
$x_\pom$ of its longitudinal momentum, with $t$ being the invariant 
momentum transfer. The parton, measured at a scale $\mu$, carries a 
fraction $\xi$ of the longitudinal momentum of the proton, or a fraction 
$\beta = \xi/x_\pom$ of the total longitudinal momentum transferred from 
the hadron. 

The experimental results for the diffractive deeply inelastic 
structure function  $F_2^{\rm diff}$ 
 are
related to the diffractive parton distributions by~\cite{bere,proo}
\begin{equation} 
\label{fact} 
F_2^{\rm diff} 
(x , Q^2 , x_\pom , t) 
= \sum_a \, \int  
d \xi   \, { df_{a/A}^{\rm diff}(\xi,x_\pom,t,\mu)
\over dx_\pom\, dt}
 {\hat F}_{ a} (x / \xi , Q^2 / \mu^2 ) 
\hspace*{0.3 cm} .
\end{equation} 
Here the hard scattering function ${\hat F}_a$ is the same as in ordinary
deeply inelastic scattering. Note that the contribution from the
diffractive gluon distribution can be significant in this formula. Although
$\hat F_{g}$ is of order $\alpha_s$, while $\hat F_{q}$ begins at order
$\alpha_s^0$, one may expect that diffractive scattering is related to
gluon exchange, so that  $f_{g/A}^{\rm diff}\gg f_{q/A}^{\rm diff}$. Note
also that the diffractive parton distribution functions depend on a
factorization scale $\mu$ (which is typically set equal to $Q$). The
$\mu$ dependence of $f^{\rm diff}$ follows from the  $\mu$ dependence
of $\hat F$. That is, the diffractive parton distributions obey the
ordinary perturbative DGLAP evolution equation. Since $\hat F$ is known
and the evolution is known, the diffractive parton distributions at a
starting scale $\mu_0$ can be determined from the experimental results~\cite{partanal,gest}.

The diffractive parton distributions at $\mu_0$ are not perturbatively
calculable. Nevertheless, one would like to have some theoretically
based insight into their behavior. To this end, notice that the problem
lies with the large transverse size of the proton. The diffractive
parton distributions for a physically small state would, in principle,
be perturbatively calculable. Quarkonium, for instance, would do. Let
us consider a slightly simpler model. Replace the proton by a special
photon that couples only to a heavy quark of mass $M \gg 1\ {\rm GeV}$
with a $\gamma^\mu$ coupling. Then the diffractive gluon distribution
can be represented in terms of cut Feynman diagrams such as the diagram
shown in Fig.~1. The top part of this diagram represents the operator that
defines  the $\overline {\rm {MS}}$ gluon distribution, as specified in
more detail below. The lower part represents the coupling of the  photon
to the heavy $Q\bar Q$ state, which couples again to the diffractively
scattered  photon in the final state. This is but one of many relevant
diagrams.

In this paper, we investigate this model for the diffractive gluon and
quark distributions, taking the limit $x_\pom \to 0$ in which ``pomeron
exchange'' dominates. Diagrams of order $\alpha_s^4$ (as in Fig.~1) are the
lowest order diagrams that make leading contributions, proportional to
$(1/x_\pom)^{2}$ as $x_\pom \to 0$. We find that we can take the limit
$x_\pom \to 0$ inside the integrals that represent the Feynman diagrams,
and express the sum of the contributing diagrams as a simple integral that
we can evaluate by numerical integration. We thus obtain an answer for the
diffractive gluon and quark distribution functions at a starting scale
$\mu_0 \approx M$.

\begin{figure}[htb]
\centerline{ \desepsf(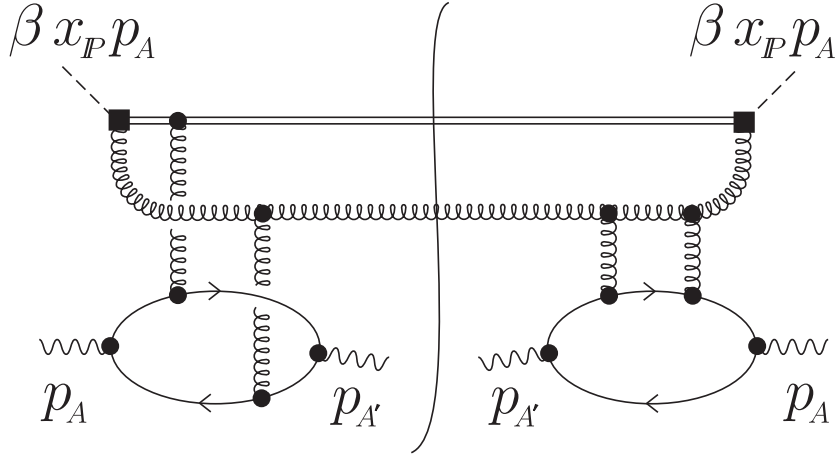 width 11 cm) } 
\caption{ A typical Feynman graph contributing to the 
diffractive deeply inelastic scattering of the model vector meson.  }
\label{figdiffpho}
\end{figure}

Evidently, the heavy quark state in the model vector meson is not the
same as the light quark state in a proton. Nevertheless, since the model
is based on nonabelian gauge theory, the qualitative features of the
resulting diffractive parton distributions may give us hints about the
real world. After giving the formulas that express the solution of the
model, we explore these qualitative features.

Consider  diffractive deeply inelastic  scattering of a hadron $A$.     
Let $(p_A , s_A )$ and 
$( p_{\!A^\prime}, s_{\!A^\prime} )$ denote the momentum and spin  
of the incident and 
the diffracted hadron.  
Let us work  
in a frame in 
which  the incident hadron is boosted along the positive 
light-cone  direction and let  $p_A^+$ be its 
`plus' 
momentum.    
The four-momentum transfer  
$q \equiv  p_A - p_{\!A^\prime}$ 
 has components  
$q^\mu = ( q^+ , q^- , {\mbox{\bf q}} )$.  
(Note that in our notation $q$ is not the virtual photon momentum,   
 as is usual in the literature.)
 The diffracted hadron can be 
characterized by 
 the fractional loss of longitudinal momentum 
$x_\pom = q^+ /  p_A^+$ 
and the invariant momentum 
transfer $ t =   ( p_A - p_{\!A^\prime} )^2 $. 
 For small $x_\pom$, $t$  is approximately 
given by $ t \simeq - {\mbox{\bf q}}^2$.

  Consider the definition of the 
diffractive parton distributions 
in terms of matrix elements of 
bilocal field
operators~\cite{bere}. This is the same  definition~\cite{cs82} as for 
inclusive parton distributions except that one requires that the final 
state include the diffractively scattered hadron. 
 For    gluons 
one has  
\begin{eqnarray}
\label{opg}
&& {{d\, f_{g/A}^{\rm {diff}}(\xi, x_\pom , t , \mu ) } \over
{dx_\pom\,dt}} = 
{1 \over {16 \, \pi^2}} \, 
{1 \over 2\pi \xi p_{\!A}^+}{1 \over 2}\sum_{s_{\!A}} \int d y^-
e^{-i\xi p_{\!A}^+ y^-}
\sum_{X,s_{\!A^\prime}}
\nonumber\\
&& \hskip 1 cm  \times 
\langle p_{\!A},s_{\!A} |\widetilde F_a(0,y^-,{\bf 0})^{+\nu}
| p_{\!A^\prime},s_{\!A^\prime}; X \rangle 
\langle p_{\!A^\prime},s_{\!A^\prime}; X|
\widetilde F_a(0)_\nu^{\ +}| p_{\!A},s_{\!A} \rangle 
\hspace*{0.2 cm} ,   
\end{eqnarray}
where 
  $\widetilde F_a(0,y^-,{\bf 0})^{+\nu}$ is 
the field strength 
operator modified by multiplication by an exponential of a line
integral of the vector potential:
\begin{equation}
\label{Ftilde}
\widetilde F_a(0,y^-,{\bf 0})^{\mu\nu}
=
\left[
{\cal P}
\exp\left(
i g \int_{y^-}^\infty d x^-\, A_c^+(0,x^-,{\bf 0})\, t_c
\right)
\right]_{ab}
F_b(0,y^-,{\bf 0})^{\mu\nu} \hspace*{0.2 cm} .  
\end{equation}
The symbol ${\cal P}$ denotes path ordering of the exponential. 
The matrices $t_c$ in Eq.~(\ref{Ftilde}) 
 are the generators of the adjoint 
representation of SU(3). 
The operator product in Eq.~(\ref{opg}) has 
ultraviolet divergences. It is understood  
that these are renormalized at the scale $\mu$  using the
$\overline{\rm MS}$ prescription.

Similarly, for  quarks of type $j$ one has 
\begin{eqnarray}
\label{opq}
&& {{d\, f_{j/A}^{\rm {diff}}(\xi, x_\pom , t , \mu ) } \over
{dx_\pom\,dt}} = 
{1 \over {16 \, \pi^2}} \, 
{1 \over { 4 \, \pi}}  
{1 \over 2}\sum_{s_{\!A}}\int d y^- e^{-i\xi p_{\!A}^+ y^-}
\sum_{X,s_{\!A^\prime}} 
\nonumber\\
&& \hskip 1 cm  \times 
\langle p_{\!A},s_{\!A} |\widetilde {\overline q}_j(0,y^-,{\bf 0})
| p_{\!A^\prime},s_{\!A^\prime}; X \rangle
\gamma^+ \langle p_{\!A^\prime},s_{\!A^\prime}; X|
{\widetilde {q}}_j(0)| p_{\!A},s_{\!A} \rangle \hspace*{0.2 cm}  , 
\end{eqnarray}
where $\widetilde q_j(0,y^-,{\bf 0})$ 
  is given by
\begin{equation}
\label{qtilde}
\widetilde q_j(0,y^-,{\bf 0})
=
\left[
{\cal P}
\exp\left(
i g \int_{y^-}^\infty d x^-\, A_c^+(0,x^-,{\bf 0})\, t_c
\right)
\right]
q_j(0,y^-,{\bf 0})  \hspace*{0.2 cm}   , 
\end{equation}
with 
$t_c$ 
 being the
generators of the  fundamental  representation of SU(3).

Consider now the case in which  $A$ is 
a photon 
 coupling  to heavy quarks,  as discussed above.  
Let us consider  small longitudinal momentum transfers, 
 $x_\pom \ll 1$, and let us 
examine   leading-power contributions
in  $x_\pom $. 
 These  
 contributions first appear in perturbation theory 
 in graphs 
  of the  same order as the one shown in Fig.~1. 
  We 
study the full set of graphs  at  this order. 
   To 
  perform this study, we use   the lightcone gauge $A^- = 0$. 
   A discussion of the gauge choice and the details of our 
  analysis will be given elsewhere~\cite{prep}. Here we 
  limit ourselves to 
 reporting the main results and discussing   their 
  implications.     
   To the leading power in $x_\pom$ we find that
the matrix elements (\ref{opg}),(\ref{opq}) 
can be written in the following form ($a = q , g$): 
\begin{eqnarray} 
\label{conv} 
&& {{d f_{a/A}^{\rm {diff}} 
(\beta,  x_\pom , {\mbox{\bf q}}^2 , M) } \over
{dx_\pom\,dt}} = 
{1 \over {64 \, \pi^2}} \,  
{1 \over 2} \, \sum_{\varepsilon} \, \sum_{\varepsilon^\prime} \, 
\int \, { {d^2  {\mbox{\bf s}}}  \over {(2 \, \pi)^2} }\,
 {1 \over { {{\mbox{\bf s}}}^2 \, 
 ( {\mbox{\bf q}}+{\mbox{\bf s}})^2 } } \,  
\int \, { {d^2  {\mbox{\bf s}}^{\prime}}  \over {(2 \, \pi)^2} }\,
 {1 \over { {{\mbox{\bf s}}}^{\prime \, 2} 
( {{\mbox{\bf q}}}+{{\mbox{\bf s}}}^{\prime })^2} } \, 
\nonumber\\ 
&& \hskip 4 cm \times 
L({\mbox{\bf q}}, {\mbox{\bf s}}, M
, \varepsilon, \varepsilon^\prime
) \, 
U_a(x_\pom ,  \beta, {\mbox{\bf q}},
{\mbox{\bf s}},{\mbox{\bf s}}^\prime) \, 
L( {\mbox{\bf q}}, {\mbox{\bf s}}^{\prime }, M
, \varepsilon, \varepsilon^\prime
) \;\;\;\;.      
\end{eqnarray} 
This result   
can be interpreted in terms of 
 graphs with  exchange of two gluons in a color-singlet state  
(such as the one depicted in Fig.~1).     
 In  $A^- = 0$   gauge,  
 only such graphs  
 contribute leading power terms as $x_\pom \to 0$.  
The denominators in the first line of 
Eq.~(\ref{conv}) come from the gluon propagators. The factor  
$U$ comes from the 
color-singlet projection of the 
two-gluon irreducible amplitude 
associated with the insertion of the operators in 
Eqs.~(\ref{opg}),(\ref{opq}). The factors $L$ describe the coupling 
of the two gluons to the incoming  quark-antiquark system. 

More precisely, the functions $U_a$ 
have the form 
\begin{equation}
\label{capu}
U_a( x_\pom , \beta , {\mbox{\bf q}},{\mbox{\bf s}},
{\mbox{\bf s}}^\prime   ) = 
{ { g_s^4 \, c_a } \over { 4 \, \pi \,  \beta \, 
(1- \beta) \, 
x_{\pom}^2 }} \, \int \,{{ d^2 {\mbox{\bf k}} } \over 
{( 2 \, \pi )^2}} \, {\mbox{Tr}} \left( u_{a}^{\dagger} 
( \beta , {\mbox{\bf k}} , {\mbox{\bf q}},{\mbox{\bf s}}^\prime )
\,  
u_{a} ( \beta , {\mbox{\bf k}} , {\mbox{\bf q}},{\mbox{\bf s}} ) 
\right) \hspace*{0.2 cm} . 
\end{equation} 
Here $c_a$ is the color factor 
and is  given for quarks and   gluons by 
\begin{equation}
\label{cqg} 
 c_q = C_F^2 \, N_c 
\hspace*{0.4 cm} , \hspace*{0.6 cm} 
  c_g = C_A^2 \, (N_c^2 - 1) 
\hspace*{0.2 cm} . 
\end{equation} 
The function $ u_{a}$ can be written as  
\begin{equation}
\label{comb}
u_{a} = 
\psi_a ( {\mbox{\bf k}} , {\mbox{\bf k}} ) 
- \psi_a ( {\mbox{\bf k}} , {\mbox{\bf k}} + 
{\mbox{\bf s}} ) 
+ \psi_a ( {\mbox{\bf k}} , {\mbox{\bf k}} - 
{\mbox{\bf q}}) - \psi_a ( {\mbox{\bf k}} , {\mbox{\bf k}} - 
{\mbox{\bf q}} - {\mbox{\bf s}} ) \hspace*{0.2 cm} .  
\end{equation} 
For $a = q$, the function 
$\psi$ has the following  
expression in terms of the $2 \times 2$  Pauli $\sigma$  matrices:    
\begin{equation}
\label{psiq}
 \psi_q( {\mbox{\bf k}} , 
 {\mbox{\bf p}} ) = { 
{\sqrt{ \beta \, (1-\beta) \, {\mbox{\bf k}}^2}   } 
\over { \beta \, {\mbox{\bf k}}^2 + (1-\beta) \, {\mbox{\bf p}}^2 }}  
\, {\mbox{\bf p}} \cdot \sigma 
\hspace*{0.2 cm} .  
\end{equation}
For $a = g$, the function 
$\psi$ is expressed in terms of two transverse vector indices 
$ i , j = 1 , 2 $ 
as 
\begin{equation}
\label{psig}
 \psi_g^{ i j} ( {\mbox{\bf k}} , 
 {\mbox{\bf p}} ) = { 
 {  \beta \, {\mbox{\bf k}}^2  \, \delta^{i j }   + 2 \, ( 1 - \beta) 
 {\mbox{\bf p}}^i \, {\mbox{\bf p}}^j } 
\over { \beta \, {\mbox{\bf k}}^2 + (1-\beta) \, {\mbox{\bf p}}^2 }}  
\hspace*{0.2 cm} . 
\end{equation} 
The functions $L$ in Eq.~(\ref{conv}) are given by 
\begin{eqnarray}
\label{genl}
&& L({\mbox{\bf q}}, {\mbox{\bf s}},  M, 
\varepsilon, 
\varepsilon^\prime) = {{e^2_Q \, e^2 \, g_s^2  } 
\over {4 \, \pi} } \, 
\int \, { {d^2 {\mbox{\bf r}}} \over {(2 \, \pi)^2} } 
\int_0^1 \, d z 
\\
&& \times {\mbox{Tr}} \left\{ 
\left[ -
\Phi^\dagger (z, {\mbox{\bf r}} + {\mbox{\bf s}}
+ z \, {\mbox{\bf q}}, M,\varepsilon^\prime )  
 - 
\Phi^\dagger (z, {\mbox{\bf r}} - {\mbox{\bf s}}
- (1-z) \, {\mbox{\bf q}}, M,\varepsilon^\prime) 
\right. \right. 
\nonumber\\
&& + \left. \left. 
\Phi^\dagger (z, {\mbox{\bf r}} + z \, {\mbox{\bf q}}, 
M,\varepsilon^\prime)  
 + 
\Phi^\dagger (z, {\mbox{\bf r}} - (1-z) \, {\mbox{\bf q}}, 
M, \varepsilon^\prime)  \right] \, 
\Phi (z, {\mbox{\bf r}} , M, \varepsilon) \right\} \hspace*{0.5 cm} ,     
\nonumber 
\end{eqnarray}
where  $e_Q$ is the quark electric charge in units of 
$e = \sqrt{ 4 \, \pi \, \alpha} \,$  and 
\begin{equation}
\label{spinphi} 
\Phi (z, {\mbox{\bf k}} ,M, \varepsilon) 
=  { 1 \over 
{  \left(  {\mbox{\bf k}}^2 + M^2 \right) } } \, 
 \left[  
(1 -  z) \, \varepsilon \cdot \sigma \, {\mbox{\bf k}} \cdot \sigma 
- z \, {\mbox{\bf k}} \cdot \sigma \, \varepsilon \cdot \sigma 
 + i \, M \, \varepsilon \cdot  \sigma   \right]
\hspace*{0.3 cm} ,  
\end{equation} 
 with  
 $\varepsilon$ and $\varepsilon^\prime$ 
being the initial and final photon transverse polarizations.

In Eq.~(\ref{conv}), the Green functions $U_a$ are universal: 
as long as the 
diffracted system is small enough that lowest order perturbation theory
applies, these functions control the scattering. The dependence on the
specific diffracted system is contained in the functions $L$. 
In Ref.~\cite{prep}, 
we will utilize a selection of hadronic functions $L$. 

The explicit form of the Green function $U_a$ is given in 
Eqs.~(\ref{capu})-(\ref{psig}).  
In the framework of $x^-$-ordered perturbation theory, the function $\psi$
in Eq.~(\ref{comb}) 
can be interpreted~\cite{prep,alig,niko}  
as the wave function for an effective
quark or gluon state created by the operator in Eq.~(\ref{opq}) 
or Eq.~(\ref{opg}).
(Here the line integral of $A^+$ in 
Eqs.~(\ref{Ftilde}),(\ref{qtilde}) represents the
interaction of an effective infinite-momentum parton with the gluon
field.)  The wave function description is natural in approaches that look
at diffractive scattering in a frame in which this quark or gluon system
has large `minus' momentum~\cite{niko,buch}. 
   Note that each of the terms in Eq.~(\ref{comb}) 
would give rise to an ultraviolet-divergent integration over $\bf k$
in Eq.~(\ref{capu}), 
but that the bad behavior cancels among the terms.  This is
because ${\bf k}^2 \to \infty$ corresponds to the partons being at the
same transverse position. But since the net color of the state is zero,
the coupling of the state to gluons vanishes in this limit.

At the leading power level,  
the $x_\pom$ dependence of the diffractive parton 
distributions   is  given simply by the overall factor $1 / x_\pom^2$ 
that we have  factored from  the integral in Eq.~(\ref{capu}). 
The $\beta$ and $t$ (or  ${\mbox{\bf q}}^2$) dependences, on 
the other hand, are non-trivial and  
come from the 
factors in the 
integrand in the right-hand side of Eq.~(\ref{conv}). The only 
explicit $\beta$ dependence in this 
integrand, in particular, is contained 
in the  functions $U_a$. 
  Inspection of  
Eqs.~(\ref{capu})-(\ref{psig})  
shows that, for   $\beta \to 0$, 
 the quark and gluon Green  functions   behave respectively as 
$U_q \sim {\mbox {const.}} \,  \beta ^0 $ and 
$U_g \sim {\mbox {const.}} \, \beta^{- 1} $.     
For  $\beta \to 1$,  
the functions $U_a$ 
 have a constant behavior 
$( 1 - \beta)^0$ at finite ${\mbox{\bf q}}^2$, ${\mbox{\bf s}}^2$ and 
${\mbox{\bf s}}^{\prime \, 2}$.

We now evaluate the diffractive distributions 
 by performing numerically 
 the integrations (\ref{conv}),(\ref{capu}),(\ref{genl}).  
  In Fig.~2  we report the results 
by  plotting the $\beta$-dependence 
of the quark and gluon diffractive 
distributions for different values of ${\mbox{\bf q}}$.  
To emphasize the  regions of small $\beta $ and 
large $\beta $ 
we make a logarithmic plot in the variable $\beta/(1-\beta)$. 
    The curves in  Fig.~2
reflect  
the  behavior of the  functions  $U_a$ 
discussed above.   
 In particular, as $\beta \to 1$ 
 the   
${\mbox{\bf q}} \neq 0$ 
 distributions 
 have a constant behavior. 
The asymptotic constants, on the other hand, are  small 
compared to the values of the distributions at 
intermediate  $\beta$. Correspondingly,   
the diffractive distributions fall off 
as one approaches the small $(1-\beta)$  region.  
Note that the gluon distribution is much larger than the quark 
distribution. Roughly, the different order of magnitude is 
  accounted for 
 by the color factors in Eq.~(\ref{cqg}), $c_g / c_q = 27/2$. 

\begin{figure}[htb]
\centerline{\desepsf(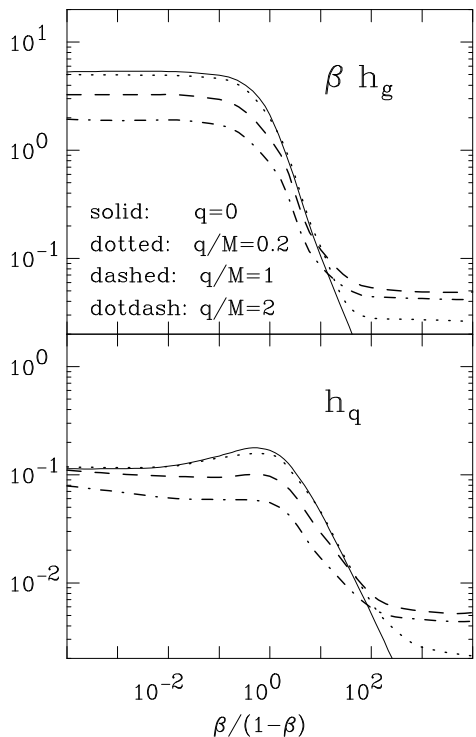 width 8 cm)}
\caption{ The $\beta$ dependence 
of the  gluon (above) and  quark (below) diffractive 
distributions for different values of 
${\mbox{\bf q}}^2 \simeq | t |$. 
 The rescaled distributions $h_a$ are defined as $ 
h_a (\beta , {\mbox{\bf q}}^2 / M^2) = 
 x_\pom^2 \, M^2 \, 
\left[ d f_{a/A}^{\rm {diff}} 
  / 
(dx_\pom\,dt) \right] / ( \alpha^2 \, e_Q^4 \, \alpha_s^4 ) $. }
\label{fighh}
\end{figure}

The calculation  described so far does not have scaling violation.  One
may  interpret the results above as a model for the diffractive
distribution functions at a scale of order $M^2$ and study the scale
dependence due to renormalization group evolution.  In Fig.~3 we report
the result of using ordinary evolution equations for the evolution of
the initial distributions up to different values of $Q^2$. In this
figure we plot the flavor-singlet quark distribution. In leading order
this is proportional to the structure function $F_2^{\rm {diff}}$.
Here we have assumed the initial scale and the mass to be $Q_0^2 = M^2
= 2 \, {\mbox {GeV}}^2$ (about the charm quark mass squared) and we have
integrated the distributions $ d f_{a/A}^{\rm {diff}}/ [dx_\pom\,dt] $
over $t \simeq - {\mbox{\bf q}}^2$ from $0$ to $M^2$.

\begin{figure}[htb]
\centerline{\desepsf(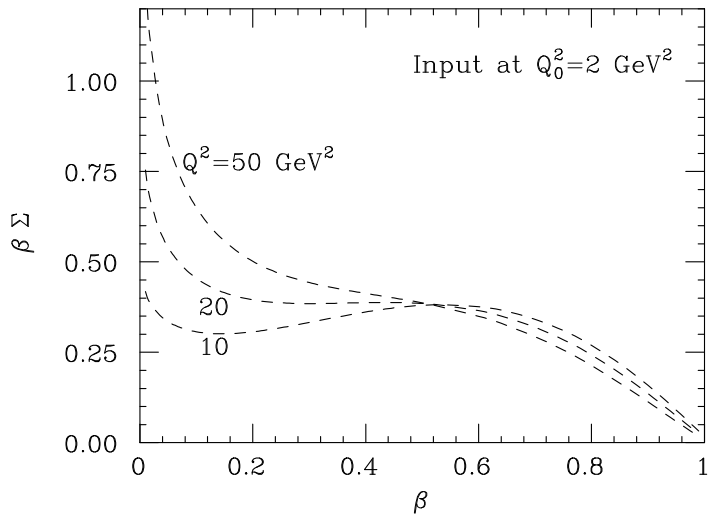 width 11 cm)} 
\caption{  $Q^2$-evolution of the diffractive singlet quark 
distribution 
\protect\linebreak 
 $  \Sigma (\beta , Q^2) 
 = {\cal N} \, \sum_{j }
\int_0^{M^2} d {\mbox{\bf q}}^2 \,  
 d f_{j/A}^{\rm {diff}} 
  / 
(dx_\pom\,dt)$, where  
$ {\cal N} = x_\pom^2 / ( \alpha^2 \, e_Q^4 \, \alpha_s^4 )$ 
and the sum runs over
$j =\{u,\bar u,d,\bar d, \dots\}$. }
\label{figsglnorm}
\end{figure}

Is there any relation between these calculations and the
data from diffractive deeply inelastic scattering at HERA?
Obviously, the protons probed in experiments at HERA have
a large transverse size, in contrast to the 
small-size hadronic state considered in our calculation.
Nevertheless, we find, by studying different hadronic wave
functions~\cite{prep}, that the $\beta$-behavior as well as the
relative size of the singlet quark and gluon distributions depends
only weakly on the specific nature of the incoming hadronic state.
This suggests that qualitative features of the results for 
small-size hadrons may apply with more generality. 

In particular, one of the most peculiar features of the 
HERA data~\cite{origdata,partanal} is
the striking difference in the $Q^2$ evolution between the diffractive
and the inclusive structure functions. In our calculation, 
 the diffractive quark distribution grows with $Q^2$ at
low $\beta$ and decreases at high $\beta$ (Fig.~3).  The stability point 
at which the behavior changes is $\beta \approx 0.5$.  This is in striking
contrast with the case of the inclusive quark distribution in a proton,
for which the stability point is at $x \approx 0.08$, but is
in qualitative agreement with the behavior found in the HERA 
experiments.  The explanation for growth in the quark distribution up
to such large values of $\beta$ in the diffractive case is that the
initial gluon distribution (Fig.~2) is large even at large values of
$\beta$. As $Q^2$ increases, the gluons feed the quark distribution
through the splitting $ g \to q {\bar q}$.

This research is supported in part by the US Department of
Energy grant DE-FG03-96ER40969. Z.K. acknowledges the 
hospitality of the Institute of Theoretical Science at the University 
of Oregon and the Theory Division at CERN while part of this work 
was being done.

\end{document}